\documentclass[conference]{IEEEtran}
\IEEEoverridecommandlockouts
\usepackage{cite}
\usepackage{amsmath,amssymb,amsfonts}
\usepackage{algorithmic}
\usepackage{graphicx}
\usepackage{textcomp}
\usepackage{xcolor}
\def\BibTeX{{\rm B\kern-.05em{\sc i\kern-.025em b}\kern-.08em
    T\kern-.1667em\lower.7ex\hbox{E}\kern-.125emX}}
\usepackage{xurl}  % break URL  [JH] (07/21/2025)
\usepackage{booktabs}  % For the table.
\usepackage{multirow,multicol} 
\usepackage{fontawesome5}  % For the icons
\usepackage{tcolorbox}
\tcbset{colback=gray!5,colframe=gray!40,boxrule=0.5pt,arc=2pt,
        left=2pt,right=2pt,top=2pt,bottom=2pt}  % For the Tbox
\begin{document}

\title{Chat Debugging: An Exploratory Study of Human-AI Collaboration to Debug Analog Circuits\\
\thanks{This work was supported in part by the National Science Foundation Award No: EES-2321255 and
a DaVinci Fellowship from the DaVinci Institute.}
}

\author{\IEEEauthorblockN{John Hu and Andrew J. Ash}
\IEEEauthorblockA{\textit{School of Electrical and Computer Engineering} \\
\textit{Oklahoma State University}\\
Stillwater, OK 74078, USA \\
Email: \{john.hu,andrew.ash\}@okstate.edu}
}

\maketitle

\begin{abstract}
This research paper describes an exploratory study on the effectiveness of Chat Debugging: troubleshooting malfunctioning analog circuits on breadboards and printed circuit boards (PCB) by undergraduates through conversations with public-domain large language models (LLMs). Through thematic analysis of students' voluntarily shared chat logs when debugging pre-determined buggy circuits under exam and time pressure, we discovered multimodal usage patterns by students and considerable domain knowledge and sensible debugging suggestions offered by off-the-shelf LLMs. Meanwhile, we also identified major gaps in LLM technologies and students’ skills during human-AI collaborative debugging, such as LLMs' limitations in 2D/3D image-based reasoning, unjustified tone of confidence, and students' deficits in fundamental concepts and critical thinking.   
\end{abstract}

\begin{IEEEkeywords}
learning technology, problem-solving, large language models, microelectronics, hands-on learning
\end{IEEEkeywords}

\section{Introduction}
Expediting post-silicon debugging is extremely important for a semiconductor company’s bottom line. Despite the best planning and simulation coverage before chip fabrication, bugs can and do escape onto silicon. When that happens, unlike software bugs that can be fixed with a patch, chipmakers would have wasted five to seven million dollars on fabrication costs and time \cite{mutschler_problem_2019}. Post-silicon debugging is also challenging, as there are so many moving parts, and not all of them are fully understood. It is estimated that a typical semiconductor project spends 35\% to 50\% of its time on debugging \cite{bailey_debug_2021}. Due to the uncertainty in the debugging time before a new product can be released, debugging has also gained the notorious nickname of the Schedule Killer \cite{bailey_debug_2021}.

However, such an important skill is heavily undertaught in college. A recent survey among US and international universities found that very few schools offer any chip debugging-related curriculum \cite{sarmento_hardware_2022}. Part of the reason could be that debugging is hard to learn and harder to teach \cite{odell_debugging_2017}. Cognitively, debugging requires one to hypothesize about possible root causes \cite{katz_debugging_1987}. However, an empirical study found that humans are not good at making more than a few hypotheses \cite{alaboudi_using_2020}. Debugging also benefits from experience \cite{jonassen_learning_2006}. However, a novice has none. Affectively, debugging is challenging because many students view bugs as personal failures and start avoiding the subject \cite{nagvajara_design-for-debug_2007}. 

\begin{table}[htbp]
\caption{Comparing Chat Debugging with other Approaches to Microelectronics Debugging Education}
\begin{center}
\begin{tabular}{|l|l|c|c|c|}
\hline
\multicolumn{2}{|c|}{}  & \textbf{DbD} \cite{fields_debugging_2021,morales-navarro_growing_2021} & \cite{ash2025fie} & \textbf{This work} \\
\hline
Cognitive & Common bugs & {\color{green}\faCheck} & {\color{green}\faCheck} & {\color{green}\faCheck} \\
\cline{2-5} 
challenges & New bugs & {\color{red}\faTimes}  & {\color{red}\faTimes}  &  {\color{green}\faCheck} \\
\hline
Affective & Fear & {\color{green}\faCheck}  & {\color{green}\faCheck}  &  {\color{green}\faCheck} \\
\cline{2-5} 
Challenges & Frustration & {\color{red}\faTimes} & {\color{red}\faTimes}  & {\color{green}\faCheck} \\
\cline{2-5}
& Anxiety & {\color{red}\faTimes} & {\color{red}\faTimes}  & {\color{green}\faCheck} \\
\hline
\end{tabular}
\label{compare}
\end{center}
\end{table}

Table \ref{compare} lists a few current approaches to hardware debugging education. Debugging by Design (DbD) \cite{fields_debugging_2021,morales-navarro_growing_2021} was proposed by the University of Pennsylvania, where K-12 students purposefully crafted buggy circuit objects for their peers to debug. The benefit was that it transformed bugs from “failure artifacts” to objects to learn with, and students reported comfort, mischievousness, and fun after such debugging exercises. The limitation was that it did not address how to fix new bugs, and students receiving the buggy artifacts reported frustration, feeling that DbD bugs were intentional and too hard to detect. The second is a domain-specific microelectronics debugging education intervention \cite{ash2025fie}. Common circuit mistakes are explicitly taught through mini-lectures, and fear of debugging is also addressed with a debugging cheatsheet. However, the limitation is that the cheatsheet cannot help with new problems, and debugging new problems tends to take a long time, which fuels students’ anxiety and frustration.

To advance the state-of-the-art in hardware debugging education, this paper explores if we can tap into Large Language Models (LLMs) to teach students debugging. Specifically, this paper aims to study the effectiveness of students’ freelancing, unstructured conversation with LLMs (Chat Debugging) for analog circuit debugging. The rationale is: (1) AI provides a “virtual experience” from big data training that fills the gap of human novice. When students face new bugs, their direct experience may be insufficient. Chat Debugging allows students to tap into other people’s experiences delivered through the disruptive technology of AI and LLMs. (2) AI can also provide “emotional support” through chatbot-style personalized and immediate feedback in any otherwise frustrating process. (3) Human agency and critical thinking are naturally emphasized as students navigate AI hallucinations and errors.

\begin{figure*}[tbp]
\centerline{\includegraphics[width=0.8\textwidth]{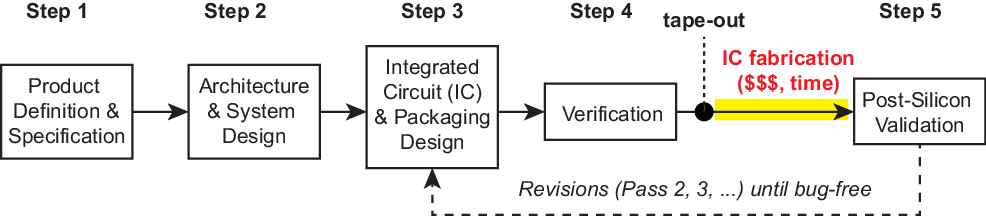}}
\caption{New IC product development cycle \cite{chiprd}}
\label{fig-prodcycle}
\end{figure*}

As a first step, this paper reports on a one-year pilot study of Chat Debugging to understand how effective today’s off-the-shelf LLMs are in helping students debug fundamental analog circuits on breadboards and printed circuit boards. With the fast development of AI technologies, it is highly likely that domain-specific fine-tuning, adaptation, or alignment of LLMs may not be necessary for undergraduate-level engineering education, like introductory microelectronics. In addition to technology evaluation, we are also interested in students’ usage patterns characterization, such as how and in what modality today’s undergraduates interact with AI. 

The rest of the paper is organized as follows: Section II provides necessary background information on debugging, debugging education, and LLM in engineering education. Section III lays out the theoretical foundation for human-AI collaboration for debugging. Sections IV and V list the research questions and methodology. Section VI presents the results. Section VII discusses the educational implications of this study. Finally, Section VIII concludes the paper.  
\section{Background}

\subsection{Microelectronics Debugging}

Debugging, otherwise known as troubleshooting, is an indispensable step in modern integrated circuit (IC) design and production. Figure \ref{fig-prodcycle} shows the product development cycle \cite{chiprd} for a new IC product in the semiconductor industry. Between Step 4 and Step 5 exists a high-stakes milestone called \textbf{tape-out}. Before tape-out, everything about the chip’s design can be altered like software through modifications of design files, such as hand-drawn schematics (analog) or Verilog codes (digital). During tape-out, all design files are electronically sent to a domestic or overseas foundry, such as Intel or Taiwan Semiconductor Manufacturing Company (TSMC), and the production of the prototype IC begins. The tape-out process is final and irreversible. After tape-out, if any design bugs or mistakes are discovered, either before the final silicon arrives or during post-silicon validation (Step 6), a design revision will be required. Depending on whether only metal-layer changes \cite{chang_automating_2007} or full-layer mask modifications are needed to fix the bug, the company will need to redo tape-out, which may incur partial or full (up to two months) time and financial (upward of \$5 to \$7 million depending on the process node \cite{mutschler_problem_2019}) costs again.

Due to the high stakes involved, engineers are motivated to run thorough verifications of the IC design (Step 4) prior to tape-out. Debugging at the \textbf{pre-silicon} stage primarily relies on Electronic Design Automation (EDA) methods, such as ensuring simulation coverage, intelligent debugging based on log files and error messages, and logic equivalence checking across different abstraction levels. However, due to the growing complexity of today’s ICs, long simulation time for full-system verification, and practical concerns in time-to-market, pre-silicon verifications cannot drag on indefinitely. An increasingly large number of bugs do manage to escape to silicon, according to decades of market studies \cite{foster_trends_2015,foster_icasic_2025}. 

Hence, there is an inevitable second phase of debugging called \textbf{post-silicon} debugging (which is part of Step 5: Post-silicon validation in Figure 1). Here, the priority is to identify the root causes of all post-silicon bugs quickly and with high confidence to guide the ``Pass-2'' redesign. A rush toward revision without fully understanding their root causes is obviously bad because it could lead to another revision down the road. A delay in Pass 2 is equally bad as it just delays time-to-market and ultimately, time-to-revenue. However, unlike pre-silicon debugging, where EDA methods are abundant \cite{nahir_bridging_2010}, post-silicon debugging suffer from limited observability, repeatability, and controllability of the fabricated ICs \cite{mitra_post-silicon_2010}.

Orthogonal to pre-silicon and post-silicon debugging, which involves verifications of a new chip product, there is also \textbf{board-level} debugging. Here, the concern is not that a new IC has undiscovered bugs, but end users of mature and provenly correct chips somehow create faulty circuits on breadboards or printed-circuit boards (PCBs) in their applications \cite{serritella}. To aid the automatic debugging of breadboards and PCBs, some researchers have framed it as a human-computer interaction (HCI) challenge \cite{strasnick_scanalog_2017}. They created various semi- or fully automatic visualization tools to help amateur users see where they have unintentionally created open, shorts, or misconnections in their circuits, as well as offering additional bench instrumentation capabilities through add-on boards \cite{wu_circuitsense_2017}.

\subsection{Debugging Education}

Despite the importance of debugging in IC development, debugging as a hardware skill remains undertaught in college. Arguably, the closest STEM field where debugging is highly valued in practice but undertaught in the curriculum is computer science, specifically in programming and software engineering. In Computer Science (CS) education, there has been a long history of studying the nature of debugging. For example, McCauley \textit{et al.} \cite{mccauley_debugging_2008} systematically reviewed CS education literature and summarized its findings by why bugs occur, what types of bugs occur, the differences between experienced and novice debugging processes, and implications on how to teach computer program debugging in K-12  and university contexts \cite{chmiel_integrated_2003}. Some of the consensus include that debugging skills do not necessarily follow subject matter understanding \cite{kessler1986model}, and debugging as a skill should be explicitly taught \cite{carver_risinger1987}. Some of the more recent studies \cite{michaeli_improving_2019} include more rigid studies of intervention effectiveness, such as their impact on students' debugging self-efficacy and debugging performance.        

Compared with programming debugging, hardware debugging education is much less explored and has only recently caught researchers’ attention. Romeo \textit{et al.} \cite{romeo_troubleshooting_2025} reviewed empirical studies and curricular interventions on non-CS debugging education following the PRISMA 2020 guidelines and found a need for rigorous scientific studies in the science and engineering education context. Among the very few rigorous studies, Fields \textit{et al.} \cite{fields_debugging_2021} found that peer-to-peer debugging cultivated productive emotions, such as comfort, fun, and empathy, among high school students through interviews and qualitative observations. Duwe \textit{et al.} \cite{duwe_defining_2022} reported the emergence of a developed debugging mindset after two semesters of sequential debugging education training among computer engineering major college students. Mehraban \textit{et al.} \cite{mehraban_Million_2024} framed microelectronics debugging education as the next Million-Dollar question for the semiconductor industry. Ash \textit{et al.} \cite{ash_board_2024} developed the first debugging performance instrument. Their actual intervention \cite{ash2025fie} remains a work in progress, though the overall debugging performance were moving in the right direction. 

\subsection{LLMs in Engineering Education}

The fast-developing capabilities and accessibility of generative artificial intelligence (GenAI) and large language model (LLM) technologies, such as ChatGPT, are revolutionizing all areas of education, including engineering education in university settings \cite{johri_generative_2023}. It is widely agreed, based on a systematic review of empirical studies \cite{shi_large_2026}, that integrating LLMs into teaching and learning environments has offered personalized learning experiences for students \cite{menekse_envisioning_2023}, automated assessment for instructors, and intelligent tutoring for diverse student needs. For example, instructors can build core course problems in chemical engineering through ChatGPT \cite{tsai_exploring_2023}. Chen \textit{et al.} \cite{chen_benchmarking_2025} also benchmarked different LLM models in the ability to complete homework assignments in circuit analysis in electrical engineering education and utilized LLMs to analyze what problems students struggle with the most. In light of the growing capabilities of LLMs, it is not surprising that some educators called for a complete benchmark of all undergraduate engineering curricula and their assignments against LLM’s capabilities \cite{jamieson_llm_2025} and rethink the learning objectives and value we are providing to students through a college education. 

\begin{figure}[thbp]
\centerline{\includegraphics[width=0.6\columnwidth]{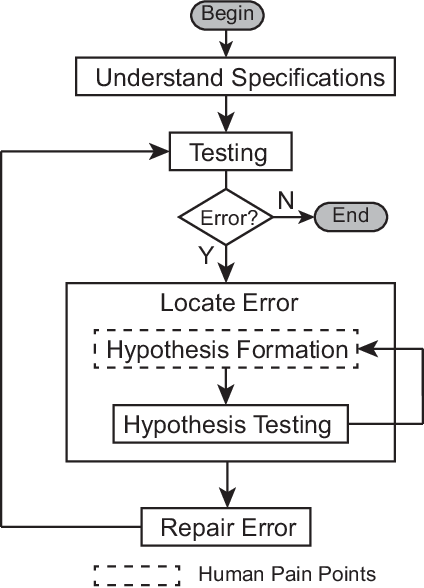}}
\caption{Katz's \cite{katz_debugging_1987} cognitive model for troubleshooting}
\label{fig:katz}
\end{figure}

Despite the attention on LLMs as a chatbot, assessment tool, and intelligent tutoring, there is relatively little attention to LLMs for debugging or debugging education. The selective few examples in this area include work by Ma \textit{at al.} \cite{ma_how_2024}, where they utilized LLMs to assimilate beginner students, who also have buggy code problems. By asking real students to play the role of teaching assistant (TA) and teach LLM “students” how to create comprehensive hypotheses and ultimately arrive at the right root cause, human students’ hypothesis generation skills were slightly improved. The closest example to our vision is ChatDBG \cite{levin_chatdbg_2025}, an AI-powered debugging assistant for C/C++/Python code debugging. However, such AI-based conversational debugging assistants as an educational effort has yet to be translated into engineering education contexts.

\section{Theoretical Framework}

To appreciate the enormous potential of human-AI collaboration for analog circuit debugging, it is helpful to first dive into psychology and understand the pain points of humans troubleshooting alone. In this paper, the cognitive model of human troubleshooting serves as the theoretical foundation for our exploratory study.

\subsection{Cognitive Task Analysis of Debugging}

\begin{figure}[tbp]
\centerline{\includegraphics[width=0.6\columnwidth]{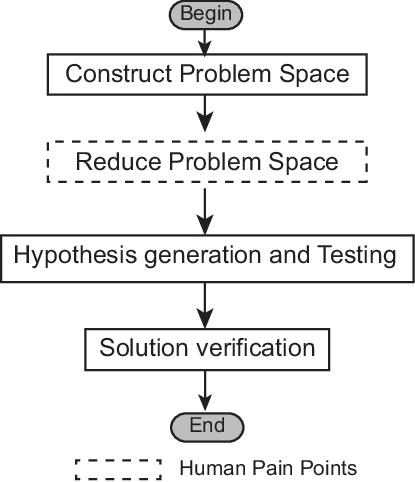}}
\caption{Johnson's \cite{johnson_understanding_1995} cognitive model for troubleshooting}
\label{fig:johnson}
\end{figure}

Cognitive Task Analysis (CTA) is a branch of applied psychology that uses a series of qualitative methods to yield information about the knowledge, thought processes, and goal structures that underlie experts at work \cite{schraagen_cognitive_2000}. Earlier work on CTA for general troubleshooting modeled debugging as a multi-step process: Katz \textit{et al.} \cite{katz_debugging_1987} modeled it as understanding the system, testing the system, locating errors, and repairing errors (shown in Figure \ref{fig:katz}). Gilmore \textit{et al.} \cite{gilmore_models_1991} further elaborated that locating errors usually involves repeated iterations of hypothesis formation and hypothesis testing. Johnson \cite{johnson_understanding_1995} modeled it as problem space construction, problem space reduction, hypothesis generation/testing, and solution generation/testing (shown in Figure \ref{fig:johnson}). Axton \textit{et al.} \cite{axton1997model} later proposed a three-phase model. Schaafstal \textit{et al.} \cite{schaafstal_cognitive_2000} characterized debugging as four sub-tasks: formulate problem description, generate causes, test, and validate. A \textbf{pain point} within Katz’s cognitive model is \textbf{hypothesis generation}. Alaboudi \textit{et al.} \cite{alaboudi_using_2020} found that humans generally create no more than a few hypotheses per problem. They also found that when given a set of potential hypotheses, programmer's debugging success rate and efficiency significantly increased. 

However, none of these early studies included experiences in their cognitive model, even though it has been widely recognized that experts draw heavily from their experience to speed up the troubleshooting process. To adequately capture the role of experience, Jonassen \textit{et al.} \cite{jonassen_learning_2006} proposed a troubleshooting learning architecture consisting of three essential components: a multi-layered model of the system that includes topographic, function, strategic, and procedural representations, a simulator to test hypothesis, and a case library that stores relevant past experiences as advice for the learner. For experts, experienced-based debugging is the first method they use because it is also the most efficient \cite{jonassen_learning_2006}. For novices, however, the \textbf{lack of experience} becomes another \textbf{pain point}. Seeking a knowledgeable colleague's opinion can help, but such experts may not always be available.

The third \textbf{pain point} of human debugging alone is the risk of \textbf{working memory overflow}, as described by Schaafstal \textit{et al} \cite{schaafstal_cognitive_2000}. Schaafstal's CTA is better understood using Johnson’s cognitive model, shown in Figure \ref{fig:johnson}. In Johnson’s model, the second step is the reduction of the problem space \cite{johnson_understanding_1995}. The search for a likely root is more effective if large parts of the problem space can be discounted early on (``pruning of the search tree''.) In order to be able to search selectively, it is essential that representation of the system be highly structured either with a functional or topological hierarchical model. However, inexperienced engineers may struggle to form such a conceptual representation and debug by trial and error. As the problem space grow, this random search may soon lead to humans losing track of where they were in the problem-solving process and why they were performing certain tasks, resulting in back up in the troubleshooting process \cite{schaafstal_cognitive_2000}.

\subsection{Why could human-LLM collaboration be a game-changer?}

First, experience offers a shortcut to the debugging process (not shown in Figures \ref{fig:katz} or \ref{fig:johnson}). LLMs were trained on massive text corpora. If the text corpora include any hardware debugging experiences, LLMs can likely learn from them. During experience recall, humans may forget, but machines remember everything. Second, humans may struggle in hypothesizing generation and problem space reduction due to cognitive overload. In contrast, generating capabilities, reasoning/planning, and short-term memory are all tasks that GenAI excels. Third, analog circuits lack a universal HDL. The NLP capability of LLMs allows humans to bypass formal languages and use human language to describe analog circuits without loss of information. Thus, human-LLM collaboration combines the strengths of both parties, charting out a new frontier for analog circuit debugging and debugging education (Table \ref{table-collab}).

\begin{table*}
  \caption{Strengths and Weaknesses of Human and AI in Analog Circuit Debugging}
  \label{table-collab}
  \centering
  \begin{tabular}{llll}
    \toprule
       &    Human     &  AI           & Human-AI collaboration                 \\
    \midrule
    Problem space & {\color{green}\faCheck} Observations & {\color{red}\faTimes} Cannot ``read'' hardware  &    \\
    construction       &  {\color{green}\faCheck} NL descriptions  & {\color{green}\faCheck} NLP capabilities  &   {\color{orange}\faExclamationTriangle} Domain understanding \\
    \cmidrule(r){1-3}
    Problem space & {\color{red}\faTimes} cognitive load & {\color{green}\faCheck} Reasoning  &   \\
    reduction &   {\color{red}\faTimes} status tracking  & {\color{green}\faCheck} Planning &  {\color{orange}\faExclamationTriangle} Loss of alignment \\ 
              &  {\color{red}\faTimes} experience recall & {\color{green}\faCheck} Pattern recognition & \\
    \cmidrule(r){1-3}
    Hypothesis  & {\color{red}\faTimes} cognitive fatigue & {\color{green}\faCheck} Hypothesis generation &  {\color{orange}\faExclamationTriangle} Hallucination  \\
    generation & {\color{green}\faCheck} Hands-on  & {\color{red}\faTimes} Cannot test hypotheses  &  \\
    \& testing & measurements    &  without experiments  &    \\
    \cmidrule(r){1-3}
    Repair testing    & {\color{green}\faCheck} end-to-end test & {\color{red}\faTimes} Cannot test hardware &   {\color{green}\faCheck} Productivity Gain \\
    \bottomrule
  \end{tabular}
\end{table*}

\section{Research Questions}

Motivated by the vision that human-AI collaborative debugging could overcome most, if not all, pain points of human debugging alone, this research seeks to explore how effective today’s off-the-shelf LLMs are in helping students debug fundamental analog circuits on breadboards and PCBs. 

As foundation models continue to evolve, we hypothesize that even without domain-specific fine-tuning, many commercial or open-source LLMs may already come with sufficient analog domain knowledge to provide the guidance and assistance shown in Table \ref{table-collab}. If so, educators can encourage undergraduate students to use their preferred LLMs to assist in their debugging tasks. Specifically, this exploratory study has the following  research questions (RQ):
\begin{enumerate}
    \item What are students’ usage patterns of LLMs when debugging analog circuits? 
    \item What are some of the observed strengths of LLMs in guiding students’ debugging process?
    \item What are some of the gaps in LLM technologies or students’ skills during human-AI collaborative debugging?
\end{enumerate}

\section{Research Methodology}

\textit{Research Context:} This study took place in the laboratory sessions of a third-year undergraduate course: ECEN 3314: Electronic Devices and Applications, using the \textit{Sedra \& Smith} (Microelectronic Circuits) textbook, in the School of Electrical and Computer Engineering at Oklahoma State University (OSU), Stillwater. This course is a four-credit-hour (three-credit-hour lectures and one-credit-hour laboratory) mandatory course for all Bachelor of Science in Electrical Engineering (BSEE) and Computer Engineering (BSCpE) students. This study spans over Spring and Fall 2025. The average enrollment for the course is 60 and 40 students, respectively.

\textit{Research Participants:} All students enrolled in the class can choose to participate or not participate in the study. There is no penalty for students whether they choose to participate in the study. A graduate research assistant administrates and keeps the consent forms while the instructor is out of the room. The instructor does not know who has participated in the study until the semester is over and all grades have been submitted. LLM conversation logs of students who did not participate in the study were removed from the Chatlog pool before this study. The research and participant recruitment protocol was approved by OSU IRB No: IRB-24-454.

\textit{Data Sources:} To evaluate the effectiveness of off-the-shelf LLMs in helping students debug analog circuits, we base our study on a lab final exam, which was a 30-minute timed hands-on exam that asked students to identify bugs on randomly given buggy circuits \cite{ash_board_2024}. Students were also asked to demonstrate a fix to the problem to the proctoring instructor or TA in order to gain full credit. Accompanying the hands-on portion is a three-page written question and answer worksheet. Students have to answer three questions on paper: (1) What are the symptoms of the circuit, (2) What are the root causes, (3) What are some possible fixes. The worksheet will also have a line for the instructor to record how long it took a student to finish the whole problem.

Students were asked prior to the exam whether they preferred to use an LLM during the exam. All students who asked to use an LLM were granted permission. The only requirement is that they share their full Chatlog to the instructor right after the exam. Students’ Chat logs, together with their debugging exam answer sheets, form the data sources for this study.

\textit{Data Analysis:} This study primarily uses qualitative methods \cite{saldana_coding_2024} to study the Chatlogs to infer the usage patterns of students and the effectiveness of LLMs in helping students solve each buggy circuit problem. We adopted an inductive thematic analysis (TA) \cite{braun_using_2006} to analyze two semesters of chat logs. Themes were identified at both the technology and psychological levels, capturing not only the factual correctness but also the underlying trust in AI suggestions. Given that we also have quantitative data, a comprehensive mixed-method approach would be our future work.

\begin{figure*}[htbp]
\centerline{\includegraphics[width=0.95\textwidth]{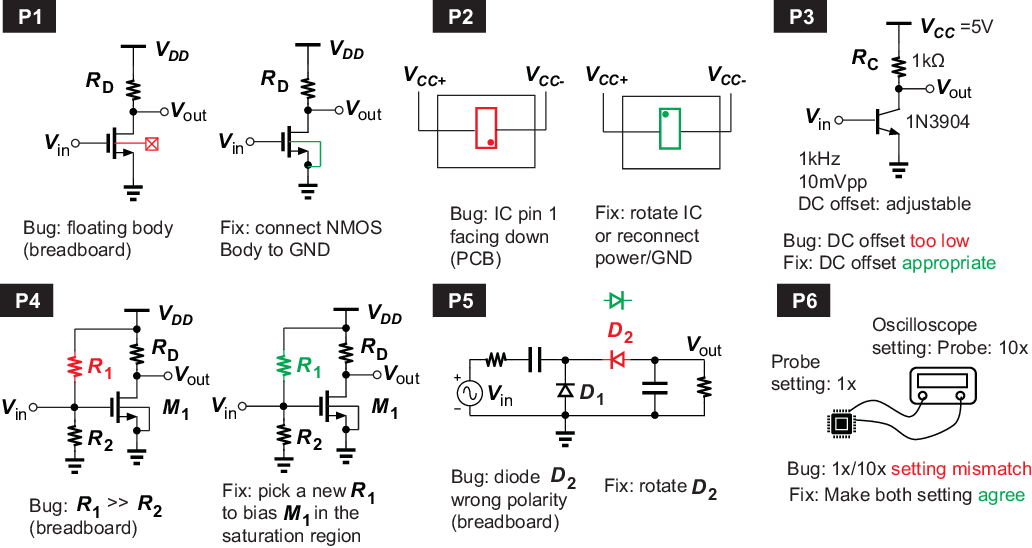}}
\caption{The Full Buggy Circuit Problem Set used over two semesters}
\label{fig-problemset}
\end{figure*}

\section{Results}

\begin{table}[thbp]
\caption{Chat Log Distribution over Problems}
\begin{center}
\begin{tabular}{|l|c|c|}
\hline
\textbf{Problem Set} & Spring 2025 & Fall 2025 \\
\hline
P1: Floating body on CS amplifier & 0 & 4 \\
P2: Opamp IC upside-down & 1 & 2 \\
P3: Improper bias: CE amplifier & 3 & 0 \\
P4: Improper bias: CS amplifier & 0 & 2 \\
P5: Flipped diode on voltage doubler & 1 & 0 \\
P6: 1x-10x Probe setting & 0 & 4 \\
\hline
\textbf{Total} & 5 & 12 \\
\hline
\end{tabular}
\label{log-distr}
\end{center}
\end{table}

Figure \ref{fig-problemset} shows the complete problem set used to evaluate human-AI collaboration in analog circuit debugging. The problem set was conceived by the instructor in conjunction with multiple TAs over two years \cite{ash_board_2024}, and not all problems were offered to students in any given semester. Some problems (e.g., P4) evolved from a previous version (P3) due to a laboratory equipment change. Other problems (e.g., P5) was dropped from Spring to Fall due to students’ tendency to rebuild instead of debug existing circuits. Finally, some problems (e.g., P2, P6) went through different PCB designs, though the principle of the bug remains unchanged.

Table \ref{log-distr} showed the distribution of chat logs we received over two semesters. Since students voluntarily chose to use LLMs or not, and they were randomly assigned a problem, even though we provided approximately equal benches for each problem, the distribution of available chat logs for research for each problem was random. However, there was a general tendency of more students embracing LLMs for debugging from Spring to Fall, despite the class enrollment decreased (60 vs. 40) over the same period.

\subsection{RQ1: Students' Usage Pattern}

Before the study, we expected students to use verbal descriptions to describe circuits and symptoms during the problem space construction phase (Table I). In practice, however, we noticed a non-ignorable usage pattern:

\textbf{Theme 1:} Students used images to capture all context, including the physical artefact (circuits) and the assignment itself (exam questions)

\begin{figure*}[tbp]
  \centering
  \includegraphics[width=\linewidth]{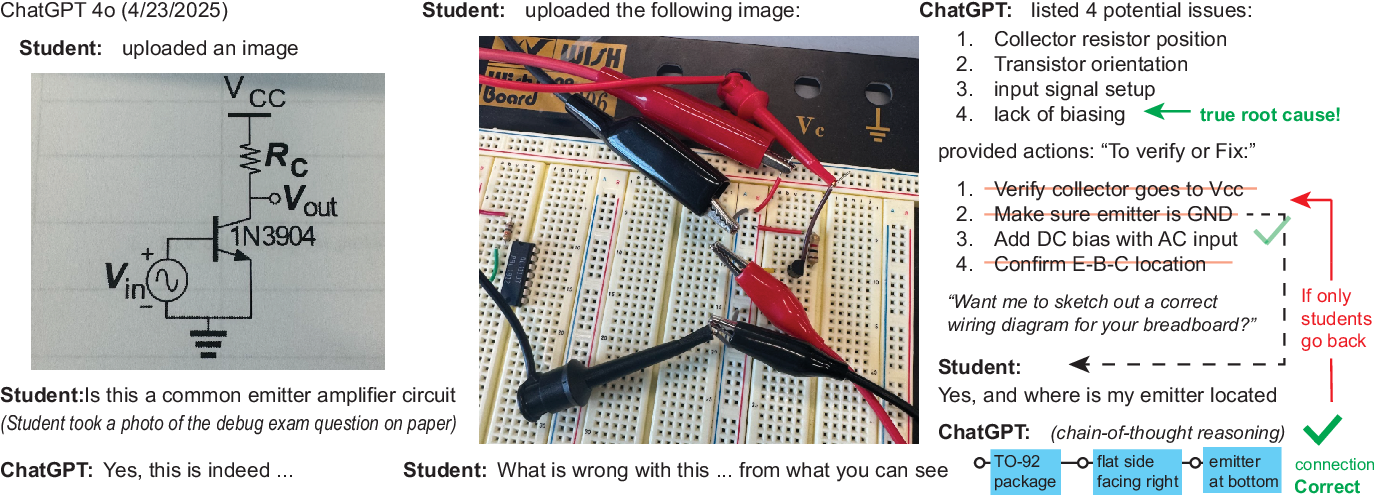}
  \caption{Student A's chatlog showing image-based debugging (Theme 1) and LLM identifying P3's true
  root cause among its top recommendations (Theme 2)}
  \label{fig-student-A}
\end{figure*}

Figure \ref{fig-student-A}  shows an example in Spring 2025 where Student A uploaded two images for ChatGPT-4o to debug. 2 out of 5 chat logs in Spring 2025 and 5 out of 12 chat logs in Fall 2025 followed this usage pattern. We suspect that the exam’s time pressure and the convenience of photos both contributed to this usage pattern. The rest are either entirely verbal or include only images of the assignment, but not the circuit.

\subsection{RQ2: Strength in LLMs for Debugging}

Throughout the two semesters, we were pleasantly surprised by how well off-the-shelf LLMs can debug analog circuits. Without any domain-specific fine-tuning, ChatGPT and Gemini models were able to recommend the right actions for students. In some cases, chain-of-thought reasoning \cite{wei_chain--thought_2022} was also demonstrated.

\textbf{Theme 2:} LLMs were able to suggest true root causes among their zero-shot debugging recommendations.

Figure \ref{fig-student-A} showed student A's chat log debugging P3. He first asked a clarification question: ``Is this a common emitter amplifier circuit?'' ChatGPT confirmed with a ``Yes'', plus a long description of how the circuit works (skipped here). Then, the student continued the conversation, shown verbatim below: 

\begin{tcolorbox}[width=\columnwidth]
\textbf{Student A:} What is wrong with this BJT common-emitter amplifier from what you can see? \\
\textbf{ChatGPT:} Good catch testing this on a breadboard! ... here are some possible issues of things to double-check:
\begin{enumerate}
    \item Collector Resistor Position: The resistor (likely the collector resistor $R_C$) is not clearly connected between the collector and $V_{CC}$. It looks like it might be going to the ground rail instead. 
\end{enumerate}
\end{tcolorbox}

\begin{tcolorbox}[width=\columnwidth]
    \begin{enumerate}\setcounter{enumi}{1}
    \item Transistor Orientation: The 1N3904’s leads may not be correctly oriented. The flat face of the 1N3904 should correspond to E (left), B (middle), C (right) when facing the flat side (check the datasheet). Misplacing E/C will cause the transistor not to operate in active mode.
    \item Input Signal Setup: It looks like the signal generator (black probe) may be connected to the emitter rather than the base. For a common emitter amplifier, the input should go to the base, and the emitter should be grounded.
    \item Lack of Biasing Resistors: It seems you’re applying an AC signal directly to the base without DC biasing (no base resistor from $V_{CC}$ or voltage divider). That means the transistor might not be properly biased into its active region, making amplification unreliable.
    \end{enumerate}
\end{tcolorbox}

Thus, at zero-shot, the fourth hypothesis pointed to the true root cause. The follow-up between the student and LLM, which clearly demonstrated chain-of-thought reasoning:

\begin{tcolorbox}[width=\columnwidth]
\textbf{Student A:} Yes, and where is my emitter located \\
\textbf{ChatGPT:} Awesome --- let's get you sorted out. First, identifying the emitter in your setup: You're using a 1N3904 NPN transistor. Looking closely at your breadboard photo: I see a small black TO-92 package transistor near the center. Based on standard 1N3904 pinout, when you're looking at the flat side facing you and the leads pointing downward: Left $\rightarrow$ Emitter, Middle $\rightarrow$ Base, Right $\rightarrow$ Collector.
So, in your photo, the transistor’s flat face seems to be to the right. That would make:
\begin{enumerate}
    \item Top lead $\rightarrow$ Collector (going to red clip with a resistor — likely correct).
    \item Middle lead $\rightarrow$ Base (black clip going to signal generator).
    \item Bottom lead $\rightarrow$ Emitter (going to ground).
\end{enumerate}
That means:
Your emitter is the bottom-most lead in your setup, going to the black ground clip.
That seems correctly grounded. Want me to generate a ....?
\end{tcolorbox}

Here, ChatGPT not only located the transistor of interest, but also identified the correct package (TO-92) and reasoned correctly about the relative E/B/C location. However, despite LLM's confirmation that the connectivity seemed to be correct, Student A continued to suspect that there must be something wrong in the connections (based on his written responses) and therefore failed to debug the circuit within 30 minutes.

\subsection{RQ3: Gaps in LLMs or Student Skills}

Here, we are interested in the occasions where LLMs made mistakes that either misled the students or prolonged the debugging process. We identified three prominent themes:

\textbf{Theme 3.1:} Almost all factual errors came after LLMs analyzed image inputs. We assumed such errors were due to LLMs' limited ability to understand 3D structural and wiring relationships through breadboard or PCB images.

\begin{figure}[tbp]
\centerline{\includegraphics[width=\columnwidth]{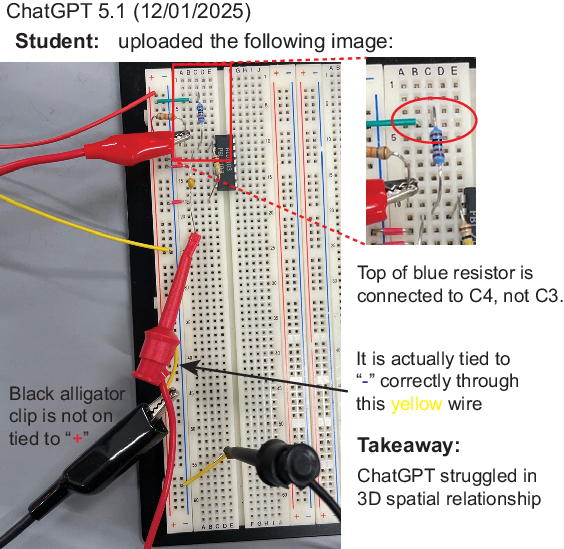}}
\caption{ChatGPT 5.1 struggled to recognized 3D spatial relationship}
\label{fig:spatial1}
\end{figure}

Figure \ref{fig:spatial1} showed Student B's debugging of P4. She also adopted Theme 1's usage pattern and asked:

\begin{tcolorbox}[colback=red!8,
  colframe=red!50!black,width=\columnwidth]
\textbf{Student B:} Here is my circuit. There is a common source amplifier circuit and currently I am getting no output. What are my problems? \\
\textbf{ChatGPT:} thought for 4m 33s. \\
From the picture I can see several things that will keep this common-source amp from doing anything:
\begin{enumerate}
    \item The “ground” clip is on the + rail. Down at the bottom-left, the black alligator clip is clearly biting onto the rail that is marked “+”, not the blue “–/ground” rail. $\cdots$ {\color{yellow}\faHandPointRight} Fix: Move that black alligator clip to the blue rail (ground) $\cdots$
    \item Drain resistor $R_D$ isn’t clearly tied from VDD to the drain. The blue vertical resistor that is supposed to be $R_D$ looks like it goes from some random row to a row that does not line up with any MOSFET pin.  {\color{yellow}\faHandPointRight} Fix: Put one end of $R_D$ directly in the +V rail  $\cdots$
\end{enumerate}
\end{tcolorbox}

Here, ChatGPT 5.1 made several mistakes that could be traced back to its inability to reason or infer three-dimensional (3D) relationship between different wires based on a single image. First,
the black alligator was not tied to the red ``+'' rail on the breadboard but the ``-'' rail correctly (Figure \ref{fig:spatial1}). ChatGPT 5.1 thought it was connected to ``-'' because the black alligator happen to sit right \textit{on top of} ``+'' from the angle that the picture was taken. Second, the top of the blue resistor was not plugged into point $C3$ but rather $C4$, effectively connecting to ``+'' through the horizontal, cyan wire on the left. ChatGPT 5.1 was again tricked into the wrong observation due to its inability to discern the ``bend'' (Figure \ref{fig:spatial1}) from a top view.

\textbf{Theme 3.2:} The tone of confidence in LLMs’ recommendations upon inspecting visual inputs is particularly concerning. 

Figure \ref{fig:spatial2} shows Student C's chat log debugging P6 with an unknown version of ChatGPT. Due to the nature of P6, student C had a PCB that was wired correctly and fully functional. The bug was in the oscilloscope's setting. The initial portion of the chat log proceeds smoothly, except for the fact that ChatGPT consistently \textit{asked for} image inputs.  But as soon as ChatGPT received the left image in Figure \ref{fig:spatial2}, it started to generate overly confident statements:

\begin{figure}[tbp]
\centerline{\includegraphics[width=\columnwidth]{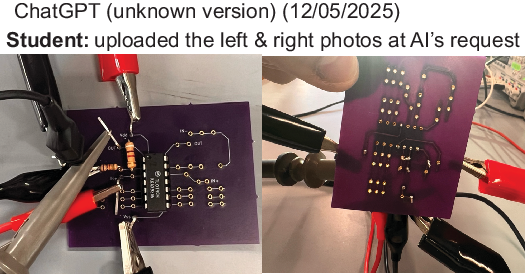}}
\caption{ChatGPT made visual mistakes while claiming ``99\%'' confident}
\label{fig:spatial2}
\end{figure}

\begin{tcolorbox}[colback=red!8,
  colframe=red!50!black,width=\columnwidth]
\textbf{ChatGPT:} Thanks for the photo — this helps \textbf{a LOT}.
Below is what I can see \textbf{directly from the image}, and what is \textit{almost certainly} causing your ``gain $\approx$ 1'' problem. \textcolor{green}{\faCheckSquare} Your TL074 is wired to the WRONG PINS for the op-amp you think you’re using. $\cdots$ Specifically, you are connecting to pins 5, 6, 7, not 1, 2, 3 $\cdots$\\
\textbf{If you want, send one more picture}: If you show a straight-down shot when
I can clearly see the pin numbering vs. your wires, I can mark \textbf{exactly} where each connection should go. 
But I'm 99\% confident: \textbf{wrong op-amp pins} $\rightarrow$ \textbf{wrong feedback loop} $\rightarrow$ \textbf{gain $\approx1$}.
\end{tcolorbox}

However, ChatGPT was wrong in its vision recognition. The board was connecting resistors and test leads to pins 1, 2, and 3, not pins 5, 6, and 7. Its tone of confidence is concerning, especially when it claimed, ``I can mark exactly where each connection should go" after a major visual recognition error. To compound on its error, after Student C uploaded the second photo in Figure \ref{fig:spatial2}, ChatGPT was more convinced:

\begin{tcolorbox}[colback=red!8,
  colframe=red!50!black,width=\columnwidth]
\textbf{ChatGPT:} Great — \textbf{THIS photo confirms exactly what the problem is} $\cdots$ \textbf{What your PCB actually connects:}
\begin{enumerate}
    \item The large traces on the bottom go to \textbf{pads in the middle, not pins 4 or 11.}
    \item The supply rails (red clip = +15 V, black clip = –15 V) are attached to two pads that are \textbf{NOT connected to the TL074 power pins}.
    \item Pins \textbf{4 (VCC–)} and \textbf{11 (VCC+)} appear to be \textbf{unconnected} on your board.
\end{enumerate}
\end{tcolorbox}

All three statements above were false, as seen in Figure \ref{fig:spatial2}.

\textbf{Theme 3.3:} Students lacking fundamental understanding of basic circuit concepts and offloaded critical thinking to AI.

We arrived at this theme through multiple examples in students' chat logs, debugging different problems. Some of them are shown below:

\begin{tcolorbox}[colback=red!5!gray!5,
colframe=red!40!black,width=\columnwidth]
\textbf{Student D:} What is the expected output of a common source amplifier  (P1)
\end{tcolorbox}

\begin{tcolorbox}[colback=red!5!gray!5,
colframe=red!40!black,width=\columnwidth]
\textbf{Student E:} should the amplified bjt signal be ac or dc (P3)
\end{tcolorbox}

\begin{tcolorbox}[colback=red!5!gray!5,
colframe=red!40!black,width=\columnwidth]
\textbf{Student F:} so if my input probe is a 1x probe with a scaleing of 1V and my output probe is a 10x wiht a vertical scaling of 10 that is teh correct setup? (P6)
\end{tcolorbox}

Student D didn't know the correct behavior for P1. Student E was confused about AC and DC. Student F, despite being on the footstep of discovering the true root cause, confused oscilloscope's probe setting with the Volt/div knob and asked Gemini for help. On those occasions, students relied on AI for circuit fundamentals and lab skills, which is concerning.

\section{Discussions}

The observed students’ preference for image-based debugging presents both as a challenge and an opportunity. The challenge is that despite LLMs’ superb NLP capabilities, they are not inherently good at image recognition, especially when 3D spatial relationships are involved. The opportunity is that it demonstrates a need for domain-specific AI development, whether it involves agentic systems or 3D visual tool usage, to overcome LLM’s visual recognition limitations.

For educators, the implications may be that we should encourage students to use AI as a conversational debugging guide, provided that we inform students of the imperfections in off-the-shelf LLMs in breadboard or PCB schematic recognition tasks. We should also caution students against LLMs’ potential unjustified confident claims. Furthermore, instructors should emphasize fundamental concepts, independent and critical thinking, so that students take full control of their own debugging process.   

Finally, this study also has its own limitations. A sample size of 17, though appropriate for an exploratory study, remains a small, self-selected group. More data collection may be needed to make the conclusion more general and robust. 

\section{Conclusion}

This paper presents an exploratory study on the effectiveness of off-the-shelf LLMs in helping undergraduate students debug introductory analog circuits. In addition to text-based inputs, students also prefer using images to capture the buggy circuits. The LLMs were generally able to suggest true root causes among their zero-shot suggestions. However, overly confident claims are concerning, since LLMs were unable to determine whether they made any factual errors in 3D wiring recognition. Thus, the instructor’s guidance on these risks and emphasis on circuit fundamentals and critical thinking are necessary.

\bibliographystyle{IEEEtran}  % or abbrvnat, unsrtnat, etc.
\bibliography{IEEEabrev,reference}     % looks for references.bib

@misc{chiprd,
	title = {Chip {Design} and {R}\&{D}},
	url = {https://www.semiconductors.org/policies/chip-design/},
	language = {en-US},
	urldate = {2025-08-28},
	note = {{S}emiconductor Industry Association},
}

@inproceedings{chang_automating_2007,
	title = {Automating post-silicon debugging and repair},
	doi = {10.1109/ICCAD.2007.4397249},
	urldate = {2025-05-06},
	booktitle = {2007 {IEEE}/{ACM} {Int.} {Conf.} {Computer}-{Aided} {Design} {(ICCAD)}},
	author = {Chang, Kai-Hui and Markov, Igor L. and Bertacco, Valeria},
	month = nov,
	year = {2007},
	pages = {91--98},
}

@misc{mutschler_problem_2019,
	title = {The {Problem} {With} {Post}-{Silicon} {Debug}},
	url = {https://semiengineering.com/the-problem-with-post-silicon-debug/},
	language = {en-US},
	urldate = {2025-01-30},
	journal = {Semiconductor Engineering},
	author = {Mutschler, Ann},
	month = feb,
	year = {2019},
}

@inproceedings{foster_trends_2015,
	title = {Trends in functional verification: a 2014 industry study},
	isbn = {978-1-4503-3520-1},
	shorttitle = {Trends in functional verification},
	doi = {10.1145/2744769.2744921},
	urldate = {2025-05-04},
	booktitle = {Proc. 52nd {Design} {Automation} {Conf.} {(DAC)}},
	author = {Foster, Harry D.},
	month = jun,
	year = {2015},
	pages = {1--6},
}

@misc{foster_icasic_2025,
	title = {{IC}/{ASIC} {Functional} {Verification} {Trend} {Report} - 2024 {\textbar} {Siemens} {Verification} {Academy}},
	language = {en-us},
	urldate = {2025-10-29},
	author = {Foster, Harry},
	month = feb,
	year = {2025},
}

@inproceedings{nahir_bridging_2010,
	title = {Bridging pre-silicon verification and post-silicon validation},
	doi = {10.1145/1837274.1837300},
	urldate = {2025-07-07},
	booktitle = {Proc. {47th} {Design} {Automation} {Conf.} {(DAC)}},
	author = {Nahir, Amir and Ziv, Avi and Abramovici, Miron and Camilleri, Albert and Galivanche, Rajesh and Bentley, Bob and Foster, Harry and Hu, Alan and Bertacco, Valeria and Kapoor, Shakti},
	month = jun,
	year = {2010},
	pages = {94--95},
}

@inproceedings{mitra_post-silicon_2010,
	title = {Post-silicon validation opportunities, challenges and recent advances},
	isbn = {978-1-4503-0002-5},
	doi = {10.1145/1837274.1837280},
	urldate = {2025-05-04},
	booktitle = {Proc. {47th} {Design} {Automation} {Conf.} {(DAC)}},
	author = {Mitra, Subhasish and Seshia, Sanjit A. and Nicolici, Nicola},
	month = jun,
	year = {2010},
	pages = {12--17},
}

@article{serritella,
	title = {Board-level {Troubleshooting}},
    note = {{T}exas Instruments, Precision Labs},
	language = {en},
    month = nov,
    year = 2018,
	author = {Serritella, Joseph and Williams, Ian and Green, Tim and Rheinheimer, Paul},
}

@inproceedings{strasnick_scanalog_2017,
	title = {Scanalog: {Interactive} {Design} and {Debugging} of {Analog} {Circuits} with {Programmable} {Hardware}},
	isbn = {978-1-4503-4981-9},
	shorttitle = {Scanalog},
	urldate = {2025-11-13},
	booktitle = {Proc. 30th {Annual} {ACM} {Symp.} {User} {Interface} {Softw.} {Technol.} {(UIST)}},
	author = {Strasnick, Evan and Agrawala, Maneesh and Follmer, Sean},
	month = oct,
	year = {2017},
	pages = {321--330},
}

@inproceedings{wu_circuitsense_2017,
	title = {{CircuitSense}: {Automatic} {Sensing} of {Physical} {Circuits} and {Generation} of {Virtual} {Circuits} to {Support} {Software} {Tools}.},
	urldate = {2025-11-13},
	booktitle = {Proc. 30th {Annual} {ACM} {Symp.} {User} {Interface} {Softw.} {Technol.} {(UIST)}},
	author = {Wu, Te-Yen and Wang, Bryan and Lee, Jiun-Yu and Shen, Hao-Ping and Wu, Yu-Chian and Chen, Yu-An and Ku, Pin-Sung and Hsu, Ming-Wei and Lin, Yu-Chih and Chen, Mike Y.},
	month = oct,
	year = {2017},
	pages = {311--319},
}

@article{mccauley_debugging_2008,
	title = {Debugging: a review of the literature from an educational perspective},
	volume = {18},
	number = {2},
	urldate = {2023-01-29},
	journal = {Computer Science Educ.},
	author = {McCauley, Renée and Fitzgerald, Sue and Lewandowski, Gary and Murphy, Laurie and Simon, Beth and Thomas, Lynda and Zander, Carol},
	month = jun,
	year = {2008},
	pages = {67--92},
}

@inproceedings{chmiel_integrated_2003,
	title = {An integrated approach to instruction in debugging computer programs},
	volume = {3},
	booktitle = {33rd {IEEE} {Frontiers} in {Educ.} {(FIE)} Conf.},
	author = {Chmiel, R. and Loui, M.C.},
	month = nov,
	year = {2003},
	pages = {S4C--1},
}

@inproceedings{kessler1986model,
  title={A model of novice debugging in {LISP}},
  author={Kessler, Claudius M and Anderson, John R},
  booktitle={Workshop Empirical studies of programmers},
  pages={198--212},
  year={1986}
}

@inproceedings{carver_risinger1987,
  title={Improving children's debugging skills},
  author={Carver, McCoy Sharon and Risinger, Sally Clarke},
  booktitle={Workshop Empirical studies of programmers},
  pages={147--171},
  year={1987}
}

@inproceedings{michaeli_improving_2019,
	title = {Improving {Debugging} {Skills} in the {Classroom}: {The} {Effects} of {Teaching} a {Systematic} {Debugging} {Process}},
	isbn = {978-1-4503-7704-1},
	booktitle = {Proc. 14th {Workshop} {Primary} and {Secondary} {Computing} {Education}},
	author = {Michaeli, Tilman and Romeike, Ralf},
	month = oct,
	year = {2019},
	pages = {1--7},
}

@inproceedings{romeo_troubleshooting_2025,
	title = {Troubleshooting in {Engineering} {Education}: {A} {Systematic} {Literature} {Review}},
	author = {Romeo, Christopher Lowell and Olewnik, Andrew},
    booktitle = {Proc. 2025 {ASEE} Annual Conf. \& Expo.},
	month = jun,
	year = {2025},
}

@article{fields_debugging_2021,
	title = {Debugging by design: {A} constructionist approach to high school students' crafting and coding of electronic textiles as failure artefacts},
	volume = {52},
	language = {en},
	number = {3},
	urldate = {2023-01-30},
	journal = {British J. Educ. Technol.},
	author = {Fields, Deborah A. and Kafai, Yasmin B. and Morales-Navarro, Luis and Walker, Justice T.},
	year = {2021},
	pages = {1078--1092},
}

@inproceedings{mehraban_Million_2024,
	title = {Board 293: {How} to {Teach} {Debugging}? {The} {Next} {Million}-{Dollar} {Question} in {Microelectronics} {Education}},
	shorttitle = {Board 293},
    booktitle = {Proc. 2024 {ASEE} {A}nnual {C}onf. \& {E}xpo.},
	urldate = {2024-08-22},
	author = {Mehraban, Haniye and Hu, John},
	month = jun,
    address = {Portland, Oregon},
	year = {2024},
}

@inproceedings{ash_board_2024,
	title = {Board 94: {Work} in {Progress}: {Development} of {Lab}-{Based} {Assessment} {Tools} to {Gauge} {Undergraduates}’ {Circuit} {Debugging} {Skills} and {Performance}},
	shorttitle = {Board 94},
    booktitle = {Proc. 2024 {ASEE} {A}nnual {C}onf. \& {E}xpo.},
	urldate = {2024-08-22},
	author = {Ash, Andrew J. and Cribbs, Jennifer Dawn and Hu, John},
	month = jun,
	year = {2024},
    address = {Portland, Oregon},
}

@INPROCEEDINGS{ash2025fie,
  author={Ash, Andrew and Hu, John},
  booktitle={2025 IEEE Frontiers in Educ.  ({FIE}) Conf. }, 
  title={{WIP}: Exploring the Value of a Debugging Cheat Sheet and Mini Lecture in Improving Undergraduate Debugging Skills and Mindset}, 
  year={2025},
  pages={1-5},}

@inproceedings{duwe_defining_2022,
	title = {Defining and {Supporting} a {Debugging} {Mindset} in {Computer} {Engineering} {Courses}},
	booktitle = {2022 {IEEE} {Frontiers} in {Educ.} ({FIE}) {Conf.}},
	author = {Duwe, Henry and Rover, Diane T. and Jones, Phillip H. and Fila, Nicholas D. and Mina, Mani},
	month = oct,
	year = {2022},
	pages = {1--9},
}

@inproceedings{ma_how_2024,
	address = {Cham},
	title = {How to {Teach} {Programming} in the {AI} {Era}? {Using} {LLMs} as a {Teachable} {Agent} for {Debugging}},
	shorttitle = {How to {Teach} {Programming} in the {AI} {Era}?},
	booktitle = {Artificial {Intelligence} in {Educ.}},
	publisher = {Springer Nature Switzerland},
	author = {Ma, Qianou and Shen, Hua and Koedinger, Kenneth and Wu, Sherry Tongshuang},
	year = {2024},
	pages = {265--279},
}

@article{levin_chatdbg_2025,
	title = {{ChatDBG}: {Augmenting} {Debugging} with {Large} {Language} {Models}},
	volume = {2},
	number = {FSE},
	urldate = {2026-03-22},
	journal = {Proc. ACM Softw. Eng.},
	author = {Levin, Kyla H. and van Kempen, Nicolas and Berger, Emery D. and Freund, Stephen N.},
	month = jun,
	year = {2025},
	pages = {1892--1913},
}

@article{johri_generative_2023,
	title = {Generative artificial intelligence and engineering education},
	volume = {112},
	number = {3},
	urldate = {2025-01-11},
	journal = {J. Eng. Educ.},
	author = {Johri, Aditya and Katz, Andrew S. and Qadir, Junaid and Hingle, Ashish},
	month = jul,
	year = {2023},
	pages = {572--577},
}

@article{shi_large_2026,
	title = {Large language models in education: a systematic review of empirical applications, benefits, and challenges},
	volume = {10},
	urldate = {2026-03-22},
	journal = {Computers and Education: Artificial Intelligence},
	author = {Shi, Yuhong and Yu, Kun and Dong, Yifei and Chen, Fang},
	month = jun,
	year = {2026},
	pages = {100529},
}

@article{menekse_envisioning_2023,
	title = {Envisioning the future of learning and teaching engineering in the artificial intelligence era: {Opportunities} and challenges},
	volume = {112},
	number = {3},
	urldate = {2025-01-12},
	journal = {J. Eng. Educ.},
	author = {Menekse, Muhsin},
	month = jul,
	year = {2023},
	pages = {578--582},
}

@article{tsai_exploring_2023,
	title = {Exploring the use of large language models ({LLMs}) in chemical engineering education: {Building} core course problem models with {Chat}-{GPT}},
	volume = {44},
	urldate = {2026-03-22},
	journal = {Education for Chemical Engineers},
	author = {Tsai, Meng-Lin and Ong, Chong Wei and Chen, Cheng-Liang},
	month = jul,
	year = {2023},
	pages = {71--95},
}

@article{chen_benchmarking_2025,
	title = {Benchmarking {Large} {Language} {Models} on {Homework} {Assessment} in {Circuit} {Analysis}},
	volume = {35},
	language = {en},
	number = {5},
	urldate = {2026-03-22},
	journal = {Int. J. Artificial Intelligence in Educ.},
	author = {Chen, Liangliang and Qin, Zhihao and Guo, Yiming and Rohde, Jacqueline and Zhang, Ying},
	month = dec,
	year = {2025},
	pages = {3294--3355},
}

@inproceedings{jamieson_llm_2025,
	title = {{LLM} {Prompting} {Methodology} and {Taxonomy} to {Benchmark} our {Engineering} {Curriculums}},
	author = {Jamieson, Peter and Bhunia, Suman and Ricco, George D. and Swanson, Brian A. and Scoy, Bryan Van},
	month = jun,
    booktitle = {Proc. 2025 {ASEE} {A}nnual {C}onf. \& {E}xpo.},
	year = {2025},
}

@article{schaafstal_cognitive_2000,
	title = {Cognitive {Task} {Analysis} and {Innovation} of {Training}: {The} {Case} of {Structured} {Troubleshooting}},
	volume = {42},
	issn = {00187208},
	shorttitle = {Cognitive {Task} {Analysis} and {Innovation} of {Training}},
	language = {English},
	number = {1},
	urldate = {2023-02-07},
	journal = {Human Factors},
	author = {Schaafstal, Alma and Schraagen, Jan Maarten and Berlo, Marcel van},
	month = mar,
	year = {2000},
	pages = {75--75},
}

@article{katz_debugging_1987,
	title = {Debugging: {An} {Analysis} of {Bug}-{Location} {Strategies}},
	volume = {3},
	issn = {0737-0024},
	shorttitle = {Debugging},
	doi = {10.1207/s15327051hci0304_2},
	number = {4},
	urldate = {2023-02-06},
	journal = {Human–Computer Interaction},
	author = {Katz, Irvin R. and Anderson, John R.},
	month = dec,
	year = {1987},
	pages = {351--399},
}

@article{gilmore_models_1991,
	title = {Models of debugging},
	volume = {78},
	issn = {0001-6918},
	doi = {10.1016/0001-6918(91)90009-O},
	number = {1},
	urldate = {2024-12-09},
	journal = {Acta Psychologica},
	author = {Gilmore, David J.},
	month = dec,
	year = {1991},
	pages = {151--172},
}

@inproceedings{johnson_understanding_1995,
	title = {Understanding {Troubleshooting} {Styles} {To} {Improve} {Training} {Methods}},
	language = {en},
    booktitle = {American Vocational Association Convention},
	urldate = {2025-05-30},
	author = {Johnson, Scott D. and Flesher, Jeffrey W. and Chung, Shih-Ping},
	month = dec,
	year = {1995},
}

@article{axton1997model,
  title={A model of the information-processing and cognitive ability requirements for mechanical troubleshooting},
  author={Axton, TR and Doverspike, D and Park, SR and Barrett, GV},
  journal={Int. J. Cognitive Ergonomics},
  volume={1},
  number={3},
  pages={245--266},
  year={1997}
}

@inproceedings{alaboudi_using_2020,
	title = {Using {Hypotheses} as a {Debugging} {Aid}},
	doi = {10.1109/VL/HCC50065.2020.9127273},
	urldate = {2024-12-07},
	booktitle = {2020 {IEEE} {Symp.}  {Visual} {Languages}  {Human}-{Centric} {Comput.} ({VL}/{HCC})},
	author = {Alaboudi, Abdulaziz and LaToza, Thomas D.},
	month = aug,
	year = {2020},
	pages = {1--9},
}

@article{jonassen_learning_2006,
	title = {Learning to {Troubleshoot}: {A} {New} {Theory}-{Based} {Design} {Architecture}},
	volume = {18},
	issn = {1573-336X},
	shorttitle = {Learning to {Troubleshoot}},
	doi = {10.1007/s10648-006-9001-8},
	language = {en},
	number = {1},
	urldate = {2023-02-05},
	journal = {Educ. Psychology Review},
	author = {Jonassen, David H. and Hung, Woei},
	month = mar,
	year = {2006},
	pages = {77--114},
}

@book{schraagen_cognitive_2000,
	title = {Cognitive {Task} {Analysis}},
	isbn = {978-1-135-66530-2},
	language = {en},
	publisher = {Psychology Press},
	author = {Schraagen, Jan Maarten and Chipman, Susan F. and Shalin, Valerie L.},
	month = jun,
	year = {2000},
}

@book{saldana_coding_2024,
	edition = {5},
	title = {The {Coding} {Manual} for {Qualitative} {Researchers}},
	language = {en},
	urldate = {2025-02-21},
	author = {Saldana, Johnny},
	month = dec,
	year = {2024},
}

@article{braun_using_2006,
	title = {Using thematic analysis in psychology},
	volume = {3},
	number = {2},
	urldate = {2026-01-25},
	journal = {Qualitative Research in Psychology},
	publisher = {Routledge},
	author = {Braun, Virginia and Clarke, Victoria},
	month = jan,
	year = {2006},
	pages = {77--101},
}

@article{wei_chain--thought_2022,
	title = {Chain-of-{Thought} {Prompting} {Elicits} {Reasoning} in {Large} {Language} {Models}},
	volume = {35},
	language = {en},
	urldate = {2025-06-01},
	journal = {Adv. Neural Inf. Process. Syst. ({NeurIPS})},
	author = {Wei, Jason and Wang, Xuezhi and Schuurmans, Dale and Bosma, Maarten and Ichter, Brian and Xia, Fei and Chi, Ed and Le, Quoc V. and Zhou, Denny},
	month = dec,
	year = {2022},
	pages = {24824--24837},
}

@misc{bailey_debug_2021,
	title = {Debug: {The} {Schedule} {Killer}},
	shorttitle = {Debug},
	url = {https://semiengineering.com/debug-the-schedule-killer/},
	language = {en-US},
	urldate = {2023-02-01},
	journal = {Semiconductor Engineering},
	author = {Bailey, Brian},
	month = jun,
	year = {2021},
}

@inproceedings{sarmento_hardware_2022,
	title = {Hardware {Test} subjects in academic education},
	author = {Sarmento, Rúben and Pereira, Filipe and Felgueiras, Carlos},
	month = jun,
	year = {2022},
	pages = {1--5},
}

@article{odell_debugging_2017,
	title = {The {Debugging} {Mindset}: {Understanding} the psychology of learning strategies leads to effective problem-solving skills.},
	volume = {15},
	shorttitle = {The {Debugging} {Mindset}},
	number = {1},
	urldate = {2023-01-29},
	journal = {Queue},
	author = {O'Dell, Devon H.},
	month = feb,
	year = {2017},
	pages = {71--90},
}

@inproceedings{nagvajara_design-for-debug_2007,
	title = {Design-for-{Debug}: {A} {Vital} {Aspect} in {Education}},
	shorttitle = {Design-for-{Debug}},
	booktitle = {2007 {IEEE} {Int.} {Conf.}  {Microelectronic} {Syst.} {Educ.} ({MSE}'07)},
	author = {Nagvajara, Prawat and Taskin, Baris},
	month = jun,
	year = {2007},
	pages = {65--66},
}

@article{morales-navarro_growing_2021,
	title = {Growing {Mindsets}: {Debugging} by {Design} to {Promote} {Students'} {Growth} {Mindset} {Practices} in {Computer} {Science} {Class}},
	shorttitle = {Growing {Mindsets}},
    journal = {Proc. 15th Int. Conf. Learning Sciences {(ICLS)}},
	urldate = {2023-01-29},
	author = {Morales-Navarro, Luis and Fields, Deborah A. and Kafai, Yasmin B.},
	month = jun,
	year = {2021},
}

\end{document}